\documentclass[aps,reprint,groupedaddress,longbibliography]{revtex4-2}
\pdfpageattr {/Group << /S /Transparency /I true /CS /DeviceRGB>>}
\usepackage{amsmath}
\usepackage{amsthm}
\usepackage{amsfonts}
\usepackage{braket}
\usepackage{graphicx}
\usepackage{wrapfig}

\begin{document}
\title{Two-channel charge-Kondo physics in graphene quantum dots}

\author{Emma L. Minarelli}
\email[]{Emma.Minarelli@ucdconnect.ie}
\affiliation{School of Physics, University College Dublin, Belfield, Dublin 4, Ireland}
\affiliation{Centre for Quantum Engineering, Science, and Technology, University College Dublin,  Dublin 4, Ireland}
\author{Jonas B. Rigo}
\email[]{Jonas.Rigo@ucdconnect.ie}
\affiliation{School of Physics, University College Dublin, Belfield, Dublin 4, Ireland}
\affiliation{Centre for Quantum Engineering, Science, and Technology, University College Dublin, Dublin 4, Ireland}
\author{Andrew K. Mitchell}
\email[]{Andrew.Mitchell@UCD.ie}
\affiliation{School of Physics, University College Dublin, Belfield, Dublin 4, Ireland}
\affiliation{Centre for Quantum Engineering, Science, and Technology, University College Dublin,  Dublin 4, Ireland}

\begin{abstract}
Nanoelectronic quantum dot devices exploiting the charge-Kondo paradigm have been established as versatile and accurate analog quantum simulators of fundamental quantum impurity models. In particular, hybrid metal-semiconductor dots connected to two metallic leads realize the two-channel Kondo (2CK) model, in which Kondo screening of the dot charge pseudospin is frustrated. In this article, a two-channel charge-Kondo device made instead from graphene components is considered, realizing a pseudogapped version of the 2CK model. The model is solved using Wilson's Numerical Renormalization Group method,  uncovering a rich phase diagram as a function of dot-lead coupling strength, channel asymmetry, and potential scattering. The complex physics of this system is explored through its thermodynamic properties, scattering T-matrix, and experimentally measurable conductance. The strong coupling pseudogap Kondo phase is found to persist in the channel-asymmetric two-channel context, while in the channel-symmetric case frustration results in a novel quantum phase transition.  Remarkably, despite the vanishing density of states in the graphene leads at low energies, a \emph{finite} linear conductance is found at zero temperature at the frustrated critical point, which is of non-Fermi liquid type. Our results suggest that the graphene charge-Kondo platform offers a unique possibility to access multichannel pseudogap Kondo physics.
\end{abstract}
\maketitle


\section{Introduction}
\noindent The Kondo effect \cite{Kondo1964} was originally discussed in the context of magnetic impurities such as Fe, in non-magnetic metallic hosts like Au. By progressively decreasing the temperature, experimental measurements revealed an unexpected resistivity minimum, attributed to enhanced electronic scattering from the impurity local moments \cite{Hewson}. The low-energy physics of such systems is explained by the deceptively simple Kondo model, which features a single spin-$\tfrac{1}{2}$ local moment exchange coupled to a featureless bath of metallic, non-interacting conduction electrons. The `Kondo effect' refers to the universal physics of this model, appearing at $T \sim T_{K}$ with $T_{K}$ an emergent low-energy scale called the Kondo temperature, in which the impurity spin is dynamically screened by a surrounding many-body entanglement cloud of conduction electrons \cite{Wilson1975}.

Since then, variants of the basic Kondo effect that arise when magnetic impurities are embedded in unconventional host materials have been studied. Examples include ferromagnets \cite{martinek2003kondo} and superconductors \cite{franke2011competition}, as well as topological materials such as topological insulators \cite{mitchell2013kondo} or Dirac/Weyl semimetals \cite{mitchell2015kondo}. However, local moments in graphene have attracted the most attention \cite{chen2011tunable,fritz2013physics}. In neutral graphene, the Dirac point is at the Fermi level \cite{neto2009electronic} and so a spin-$\tfrac{1}{2}$ impurity couples to a bath of conduction electrons with a density of states (DoS) featuring a low-energy pseudogap $\rho(\omega)\sim |\omega|^r$ with $r=1$. This has dramatic consequences for the resulting Kondo physics \cite{fritz2013physics} due to the depletion of low-energy degrees of freedom in graphene that can participate in screening the impurity spin.

Deeper insights into strongly correlated electron physics and Kondo have been gained from tunable circuit realizations of fundamental models in nanoelectronics devices, made possible by remarkable recent advances in nanofabrication and characterization techniques \cite{heinrich2021quantum,barthelemy2013quantum}. This provides a route to probing and manipulating quantum matter at the nanoscale in a way that would be impossible in bulk systems. In particular, semiconductor quantum dots (QDs) have been shown to behave like artificial atoms \cite{Kastner_Artificialatoms1994}, with the extreme quantum confinement producing a discrete level structure and strong electron-electron interactions. Coupling such quantum dots to metallic electrodes at quantum point contacts (QPCs) gives rise to the Kondo effect at low temperatures \cite{Goldhaber-Gordon-Kondo1998Exp,cronenwett1998tunable,van2000kondo}, with a single local moment trapped on the dot facilitating spin-flip scattering of lead conduction electrons. In such devices, entanglement spreads across the QD and the leads in a macroscopic `Kondo cloud' \cite{mitchell2011real,yoo2018detecting}, producing the famous Kondo resonance in electrical conductance \cite{PustilnikGlazman_review2004}. Quantum transport properties of QDs can be tuned \textit{in situ} by applying gate voltages to control the QPC transmissions and dot potential. A  bias voltage can drive the system out of equilibrium.

Quantum dot devices also allow more complex quantum impurity models to be realized experimentally by controlling the microscopic interactions. As such, they constitute a versatile platform to study a rich range of physics \cite{barthelemy2013quantum}, including quantum phase transitions (QPTs) \cite{vojta2006impurity,mitchell2009quantum}, emergent symmetries \cite{keller2014emergent,mitchell2021so}, and non-Fermi liquid (NFL) physics \cite{potok2007observation,keller2015universal,Mitchell_UniversalLowT2CK2012,mitchell2012two}. The two channel Kondo (2CK) model \cite{Nozieres_KondoRealMetal1980} is a famous example which embodies the frustration of Kondo screening of a single impurity by two distinct channels of metallic conduction electrons, and displays all these features \cite{Affleck_CriticalOverscreened_1991}. The standard 2CK model Hamiltonian reads,
\begin{equation}\label{eq:H2ck}
    \hat{H}_{\rm 2CK} = \hat{H}_{\rm leads} + J_1 \hat{\boldsymbol{S}}\cdot \hat{\boldsymbol{s}}_1 + J_2 \hat{\boldsymbol{S}}\cdot\hat{\boldsymbol{s}}_2 \;,
\end{equation}
where $\hat{H}_{\rm leads} =\sum_{\alpha\sigma\boldsymbol{k}} \epsilon_{\boldsymbol{k}}^{\phantom{\dagger
}} c_{\alpha\sigma \boldsymbol{k}}^{\dagger}c_{\alpha\sigma \boldsymbol{k}}^{\phantom{\dagger
}}$ describes two leads $\alpha=1,2$, each with spin $\sigma=\uparrow,\downarrow$ electrons with  momentum $\boldsymbol{k}$. In the original formulation of 2CK model, the dispersion $\epsilon_{\boldsymbol{k}}$ is taken to be linear at low energies such that the electronic DoS of the leads at the impurity position is flat. The metallic flat band approximation is typically employed for the free conduction electrons, $\rho(\omega)\sim \sum_{\boldsymbol{k}} \delta(\omega-\epsilon_{\boldsymbol{k}}) \equiv \rho_0 \Theta(|\omega|-D)$, describing a flat density of states $\rho_0$ inside a band of half-width $D$. In Eq.~\ref{eq:H2ck},  $\hat{\boldsymbol{S}}$ is a spin-$\tfrac{1}{2}$ operator for the impurity and $\hat{\boldsymbol{s}}_{\alpha}$ is a spin-$\tfrac{1}{2}$ operator for the spin density in lead $\alpha$ at the impurity position, such that $\hat{H}_{\rm 2CK}$ possesses SU(2) spin symmetry. The metallic 2CK model supports a QPT, with an NFL critical point at $J_1=J_2$ \cite{Affleck_CriticalOverscreened_1991}. Signatures of the critical point in this model have been observed experimentally in semiconductor quantum dot devices \cite{potok2007observation,keller2015universal}.

More recently, a new nanoelectronics paradigm has emerged, based on charge-Kondo quantum dots \cite{2CK-RenormalisationFlow_IftikharPierre2015Exp,mitchell2016universality,iftikhar2018tunable,han2021fractional,pouse2021exotic}. In the standard setup, a large QD is coupled at QPCs to source and drain leads. These devices realize anisotropic multichannel Kondo models through Matveev's mapping \cite{matveev1995coulomb,furusaki1995theory} of the macroscopic charge states of the large QD to an effective pseudospin that is flipped by electronic tunneling at the QPCs. This approach has led to unprecedented control over the frustrated 2CK state, and has uncovered the full renormalization group (RG) flow diagram through transport measurements \cite{2CK-RenormalisationFlow_IftikharPierre2015Exp}.

Motivated by these developments, in this paper we consider combining the two-channel charge-Kondo setup of Ref.~\cite{2CK-RenormalisationFlow_IftikharPierre2015Exp} with the pseudogap Kondo physics of graphene in Ref.~\cite{fritz2013physics}, to realize a novel two-channel pseudogap charge-Kondo effect. We envisage a charge-Kondo device made from graphene components (see Fig.~\ref{Fig:setup}), such that the dot charge pseudospin is coupled to two channels of conduction electrons, each with the characteristic linear pseudogap DoS of graphene. This work is a theoretical exploration of such a system and its phase diagram. We characterize the phases and phase transitions through thermodynamic quantities, and focus on experimentally-relevant physical observables such as the conductance. However, we do not claim to address the practical complexities that will inevitably arise in the experimental realization of a graphene charge-Kondo device.  

We note that the generic properties of fully spin- and channel-symmetric two-channel pseudogap Kondo models were discussed in Ref.~\cite{schneider2011two}, although the $r=1$ linear pseudogap case relevant to graphene was not analyzed in detail and a device realization was not proposed. Furthermore, our charge-Kondo implementation leads to crucial differences in the model and transport measurement setup which have not previously been considered. These differences and our new results are highlighted in the following.


\section{Model, Methods, and Observables}
\begin{figure}[h]
\includegraphics[width=\columnwidth]{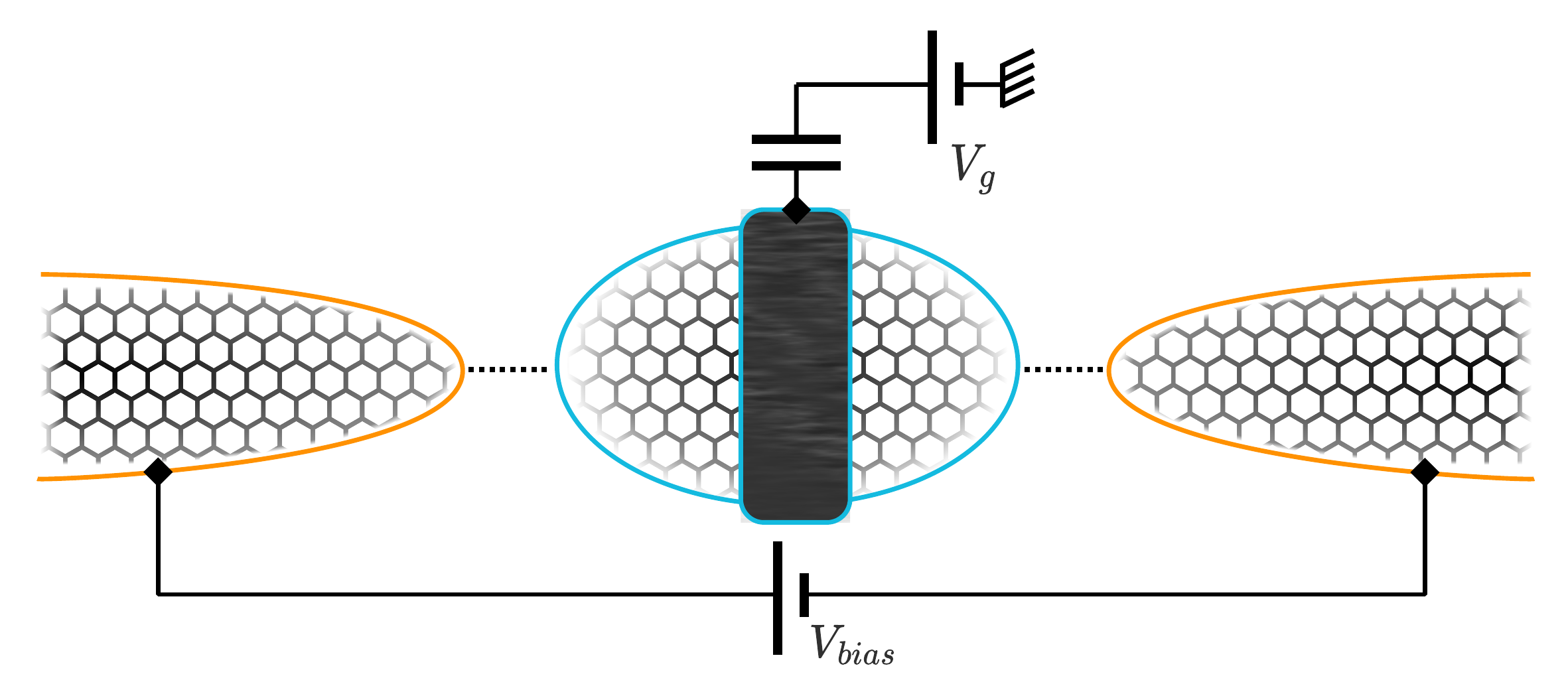}
\caption{Schematic of the two-channel graphene charge-Kondo quantum dot system. A net current flows from source to drain graphene leads through the large graphene dot in response to a bias voltage. A gate voltage $V_g$ controls the dot filling. The black bar denotes the decoherer.}
\label{Fig:setup}
\end{figure}

We consider a two-channel charge-Kondo device in which both dot and leads are made from graphene, as illustrated in Fig.~\ref{Fig:setup}. We note that graphene quantum dots have been the topic of active experimental study recently \cite{bacon2014graphene,yan2019recent,cai2019eco}. We envisage a large dot is tunnel-coupled to leads $\alpha=1,2$ at QPCs with transmission $\tau_{\alpha}$ which can be controlled \textit{in situ} by gate voltages. A plunger gate voltage $V_g$ controls the dot potential and hence the dot filling. A decoherer is interjected between the leads via an Ohmic contact on the dot (black bar in Fig.~\ref{Fig:setup}) which gives rise to a long dwell time and an effective continuum dot level spectrum (this was achieved in the experiments of Ref.~\cite{2CK-RenormalisationFlow_IftikharPierre2015Exp} using a metallic component). This results in two effectively independent electronic reservoirs around each of the two QPCs; these form the two independent channels in the 2CK model. However, tunneling events onto and off the dot are correlated by the large dot charging energy, $E_c$. The whole device is operated in a strong magnetic field so that the electrons are effectively spinless (that is, the Zeeman splitting is the largest energy scale in the problem).

\noindent The model Hamiltonian for the device illustrated in Fig.~\ref{Fig:setup} is given by,
\begin{equation}\label{Eq:Heff}
\begin{split}
\hat{H} = \hat{H}_{\rm leads} + \sum_{\alpha=1,2} ~ \sum_{\mathbf{k}\mathbf{k}^{\prime}} \Big[ &J_{\alpha}( \hat{Q}^{+} c^{\dagger}_{\alpha D\mathbf{k}}c_{\alpha L\mathbf{k}^{\prime}} + c^{\dagger}_{\alpha L\mathbf{k}^{\prime}}c_{\alpha D\mathbf{k}} \hat{Q}^{-}) \\+& W_{\alpha}(c^{\dagger}_{\alpha L\mathbf{k}}c_{\alpha L \mathbf{k}^{\prime}}+c^{\dagger}_{\alpha D\mathbf{k}}c_{\alpha D \mathbf{k}^{\prime}}) \Big ] \\ +& E_{c}(\hat{N}_{D}-N_{g})^{2} \;,
\end{split}
\end{equation}
where $\hat{H}_{\rm leads} =\sum_{\alpha\sigma\boldsymbol{k}} \epsilon_{\boldsymbol{k}}^{\phantom{\dagger
}} c_{\alpha\sigma \boldsymbol{k}}^{\dagger}c_{\alpha\sigma \boldsymbol{k}}^{\phantom{\dagger
}}$ describes the distinct conduction electron reservoirs around each QPC labelled by $\alpha=1,2$, and with $\sigma=L,D$ corresponding to lead or dot electrons (rather than physical spin $\uparrow$,$\downarrow$). For graphene components, we use the dispersion   $\epsilon_{\boldsymbol{k}}^{\pm}=\pm t\sqrt{1+4\cos^2\tfrac{a}{2}k_x + 4\cos\tfrac{a}{2}k_x\cos\tfrac{\sqrt{3}a}{2}k_y}$ for the two bands, assumed here to be independent of $\alpha$ and $\sigma$, with nearest neighbour tunneling matrix element $t\approx 2.8$ eV and lattice constant $a\approx 2.46$ \AA ~\cite{neto2009electronic}. The resulting DoS $\rho_{\alpha\sigma}(\omega)\equiv \rho(\omega)$ has a bandwidth $D=3t$ and possesses a linear pseudogap $\rho(\omega)\sim |\omega|$ for $|\omega|\ll t$.

The terms proportional to $J_{\alpha}$ describe  electronic tunneling at the QPCs between leads and dot. The tunneling matrix elements $J_{\alpha}$ are related to the bare (unrenormalized) QPC transmissions via \cite{lebanon2003coulomb} $\tau_{\alpha}(\omega)=4\pi^2\rho_{\alpha L}(\omega)\rho_{\alpha D}(\omega)J_{\alpha}^2/[1+\pi^2\rho_{\alpha L}(\omega)\rho_{\alpha D}(\omega)J_{\alpha}^2]^2$, which are in general energy-dependent for structured leads. 
States of the isolated dot with a macroscopic number of electrons $N_D$ are denoted $|N_D\rangle$, with corresponding dot number operator $\hat{N}_D=\sum_{\alpha\boldsymbol{k}}c_{\alpha D\boldsymbol{k}}^{\dagger}c_{\alpha D\boldsymbol{k}}^{\phantom{\dagger}}\equiv \sum_{N_D} N_D|N_D\rangle\langle N_D|$. Tunneling at the QPCs changes the dot charge, which we describe \cite{matveev1995coulomb} using the charge raising and lowering operators $\hat{Q}^{\pm}= \sum_{N_{D}} \ket{N_{D}\pm1}\bra{N_{D}}$. The dot has a finite charging energy that depends on the filling via the term proportional to $E_c$. The filling can be adjusted by tuning $N_g$ in Eq.~\ref{Eq:Heff}, which is controlled in experiment by the gate voltage  $V_g=V_g^0-2E_c N_g/e$. We define $\delta V_g=-2E_c (N_g-N_D^0-\tfrac{1}{2})/e$ such that the macroscopic dot charge states $|N_D^0\rangle$ and $|N_D^0+1\rangle$ are degenerate at $\delta V_g=0$. Potential scattering at the QPCs is described by the term proportional to $W_{\alpha}$.

Provided $k_{\rm B} T, e\delta V_g \ll E_c$, only the lowest two dot charge states $|N_D^0\rangle$ and $|N_D^0+1\rangle$ are accessible and relevant for transport. In this case the dot charge operators become effective pseudospin-$\tfrac{1}{2}$ operators, $\hat{Q}^+ \to \hat{S}_D^+ = |N_D^0+1\rangle\langle N_D^0|$ and $\hat{Q}^- \to \hat{S}_D^- = |N_D^0\rangle\langle N_D^0+1|$. Thus $\hat{S}_D^+$ flips the dot charge pseudospin from $\Downarrow$ to $\Uparrow$ while $\hat{S}_D^-$ flips it back. We also introduce the pseudospin operator $\hat{S}_D^z=\tfrac{1}{2}( \ket{N_{D}^0+1}\bra{N_{D}^0+1}-\ket{N_{D}^0}\bra{N_{D}^0})$. Finally, we perform a trivial relabelling $\sigma=\{ L,D\}\to \{\uparrow,\downarrow\}$ such that the electronic operators become  $c_{\alpha L \mathbf{k}} \to c_{\alpha \uparrow \mathbf{k}}$ and $c_{\alpha D \mathbf{k}} \to c_{\alpha \downarrow \mathbf{k}}$. With this, we arrive at the effective pseudogap two-channel charge-Kondo (2CCK) model studied in this paper:
\begin{equation}\label{Eq:HK}\begin{split}
\hat{H}_{2CCK} =  &\sum_{\alpha=1,2} ~ \sum_{\mathbf{k}\mathbf{k}^{\prime}} \Big [ J_{\alpha}(  \underbrace{\hat{S}_{D}^{+}c^{\dagger}_{\alpha \downarrow \mathbf{k}}c_{\alpha  \uparrow \mathbf{k}^{\prime}} }_{\hat{S}_{D}^{+}\hat{s}_{\alpha}^{-}} + \underbrace{c^{\dagger}_{\alpha \uparrow \mathbf{k}^{\prime}}c_{\alpha \downarrow \mathbf{k}} \hat{S}_{D}^{-})}_{\hat{s}_{\alpha}^{+}\hat{S}_{D}^{-}} \\&+ W_{\alpha}\sum_{\sigma}c^{\dagger}_{\alpha \sigma \mathbf{k}}c_{\alpha \sigma \mathbf{k}^{\prime}} \Big ] +e\delta V_{g}\hat{S}^{z}_{D} +\hat{H}_{\rm leads} \;.
\end{split}\end{equation}
This model is a variant of the famous 2CK model, Eq.~\ref{eq:H2ck} -- but with a few important differences. 
Firstly, the DoS of the conduction electrons described by $\hat{H}_{\rm leads}$ is not metallic, but has a low-energy $r=1$ pseudogap. 
Secondly, tunneling at the QPCs give an effective \textit{anisotropic} exchange coupling between the dot charge pseudospin and the conduction electrons. The SU(2) symmetry of Eq.~\ref{eq:H2ck} is broken in Eq.~\ref{Eq:HK} since the $z$-component of the coupling is missing. However, we find that this effective spin anisotropy is RG irrelevant in the two-channel pseudogap Kondo problem (just as for the single-channel \cite{Hewson,kogan2018poor} and two-channel \cite{Affleck_CriticalOverscreened_1991} metallic case, as well as the single-channel pseudogap case \cite{fritz2004phase}). Only the spin-flip terms are important for Kondo, and these are captured by the effective model. It should also be emphasized that the effective exchange couplings $J_{\alpha}$ originate from the QPC tunnelings; there is no underlying Anderson model, so the $J_{\alpha}$ need not be perturbatively small. In fact, since they are related to the QPC transmissions, they can become large simply by opening the QPCs \cite{2CK-RenormalisationFlow_IftikharPierre2015Exp}. This is important because Kondo physics is only realized in the pseudogap model at relatively large bare coupling strengths. 
Thirdly, the gate voltage $\delta V_g$ appears as an effective impurity magnetic field.
Finally, we have an additional potential scattering term $W_{\alpha}$. This is traditionally omitted in Eq.~\ref{eq:H2ck} because potential scattering is RG irrelevant in the metallic Kondo problem \cite{Hewson}. However, we must keep it because potential scattering is known to be important in the single-channel pseudogap Kondo model \cite{fritz2004phase}. Indeed, we find that it is strongly RG relevant in our two-channel pseudogap variant, Eq.~\ref{Eq:HK}.

Another important difference in terms of the experimental realization is the nature of the transport measurement. As illustrated in Fig.~\ref{Fig:setup}, a series current of spinless electrons is measured between the physical source and drain leads through the dot, in response to a bias voltage. But in the mapped spin model, this is an unconventional measurement: we effectively apply a bias between leads $\alpha=1,2$ but only to the $\sigma=\uparrow$ conduction electrons. Even though there is no charge current possible between leads in the original 2CK model Eq.~\ref{eq:H2ck}, the charge-Kondo setup Eq.~\ref{Eq:HK} allows an effective \textit{spin} current to be measured. 

The ac linear response electrical conductance through the device is defined,
\begin{equation}\label{Eq:Gc}
G_{C}(\omega,T) = \frac{\langle \hat{I}_{2\uparrow}\rangle}{V_{bias}}\biggr\rvert_{V_{bias}\rightarrow 0} 
\end{equation}
due to an oscillating bias described by $\hat{H}_{\rm bias}=-eV_{\textit{bias}}\cos(\omega t)\hat{N}_{1\uparrow}$ with ac frequency $\omega$. Here $\hat{I}_{\alpha \uparrow}=-e\tfrac{d}{dt}\hat{N}_{\alpha\uparrow}$ is the current operator for lead $\alpha$ (and $\sigma=\uparrow$) while $\hat{N}_{\alpha \uparrow}=\sum_{\boldsymbol{k}}c_{\alpha \uparrow \boldsymbol{k}}^{\dagger}c_{\alpha \uparrow \boldsymbol{k}}^{\phantom{\dagger}}$. We obtain the ac linear conductance from the Kubo formula \cite{IzumidaSakai1997},
\begin{equation}\label{Eq:Kubo}
\begin{aligned}
G_{C}(\omega,T) 
= \frac{-{\rm Im}\langle\langle\hat{I}_{1\uparrow};\hat{I}_{2\uparrow}\rangle\rangle_{\omega,T}}{\omega} \equiv 2\pi G_{0} \omega {\rm Im}\langle\langle\hat{N}_{1\uparrow};\hat{N}_{2\uparrow} \rangle\rangle_{\omega,T} 
 ~,
\end{aligned}
\end{equation}
where $\langle\langle\cdot~;~\cdot \rangle\rangle $
denotes a retarded real-frequency correlation function evaluated at equilibrium, and 
 $G_{0}=e^{2}/h$ is the conductance quantum ($\hbar= 1$). The second equality in Eq.~\ref{Eq:Kubo} follows from equations of motion and is found to greatly improve the accuracy of numerical calculations \cite{minarelliQT}. Note that the system is not in proportionate coupling, and so correlated electron transport coefficients cannot be expressed in terms of a Landauer-type formula \cite{MeirWingreen_1992} involving the dot spectral function.
 
 In addition to the conductance, we explore the phase diagram and RG fixed points (FPs) of the model using physical thermodynamical observables. We define the dot contribution to a thermodynamic quantity $\Omega$ at temperature $T$ as $\Omega_D(T)=\Omega(T)-\Omega^0(T)$, where $\Omega(T)$ is calculated for the full lead-dot-lead system, while $\Omega^0(T)$ is calculated only for the free conduction electrons (without the dot pseudospin). For the entropy $S_D(T)$ we use $S^{(0)}=-\partial F^{(0)}/\partial T$, with $F^{(0)}=-k_{\rm B}T\ln Z^{(0)}$ the free energy. Recently this entropy has been extracted experimentally in similar quantum dot devices by exploiting a Maxwell relation connecting the entropy change for a process to measurable changes in the dot charge \cite{child2021entropy,han2021fractional}. For the magnetic susceptibility $k_{\rm B} T \chi_D(T)$ we evaluate $k_{\rm B}T\chi^{(0)}= \braket{(\hat{S}^{z}_{tot})^{2}}^{(0)} -(\braket{\hat{S}^{z}_{tot}}^{(0)})^{2}$ at zero field ($\delta V_g=0$), with $\hat{S}^{z}_{tot}$ the $z$-projection of the total spin of the system. The role of particle-hole asymmetry will be assessed through the conduction electron `excess charge' $N_{\alpha}=\langle \hat{N}_{\alpha}\rangle - \langle \hat{N}_{\alpha}\rangle^0$ with $\hat{N}_{\alpha}=\sum_{\sigma}\hat{N}_{\alpha\sigma}$. 
 Dynamics of the system are characterized by the channel-resolved T-matrix, which describes how conduction electrons are scattered from the dot pseudospin. The T-matrix equation reads,
\begin{equation}\label{Eq:TmatEq}
\mathbb{G}_{\alpha\beta}(\omega,T) - \mathbb{G}^{0}_{\alpha\beta}(\omega) = \mathbb{G}^{0}_{\alpha\alpha}(\omega) \mathbb{T}_{\alpha\beta}(\omega,T)  \mathbb{G}^{0}_{\beta\beta}(\omega) ~,
\end{equation}
where $\mathbb{G}_{\alpha\beta}(\omega,T)$ and $\mathbb{G}_{\alpha\beta}^0(\omega)$ are, respectively, the full and free retarded electronic Green's functions at the dot position. Due to decoherence between the QPCs (resulting in separately conserved charge in each channel in Eq.~\ref{Eq:HK}), we have  $\mathbb{G}_{\alpha\beta},\mathbb{G}^0_{\alpha\beta},\mathbb{T}_{\alpha\beta} \propto \delta_{\alpha\beta}$ and the T-matrix equation becomes channel-diagonal. Furthermore, $-\tfrac{1}{\pi}{\rm Im}\mathbb{G}^0_{\alpha\alpha}(\omega)=\rho(\omega)$ is the free graphene DoS. In the following we consider the spectrum of the T-matrix for channel $\alpha$, defined as $t_{\alpha}(\omega,T)=-\tfrac{1}{\pi}{\rm Im}\mathbb{T}_{\alpha\alpha}(\omega,T)$.


\subsection{Numerical Renormalization Group}
The two-channel pseudogap charge-Kondo model, Eq.~\ref{Eq:HK}, is solved using Wilson's Numerical Renormalization Group (NRG) technique \cite{Wilson1975,Bulla2008,Weichselbaum2007}, which provides numerically-exact access to the physical quantities discussed in the previous section.

The first step is the logarithmic discretization of the conduction electron DoS, and subsequent mapping of $\hat{H}_{\rm leads}$ to Wilson chains \cite{Wilson1975,Bulla2008},
\begin{equation}\label{Eq:HBathWC}
\hat{H}_{\rm leads} \to \hat{H}^{\rm disc}_{\rm leads}=\sum_{\alpha,\sigma}~\sum_{n=0}^{\infty}t_{n}\left( f^{\dagger}_{\alpha \sigma n}f_{\alpha \sigma n+1}^{\phantom{\dagger}} + f^{\dagger}_{\alpha \sigma n+1}f_{\alpha \sigma n} ^{\phantom{\dagger}} \right) ~.
\end{equation}
The dot then couples to the end of the Wilson chains, at site $n=0$. 
The logarithmic discretization is parameterized by $\Lambda$, with the continuum description being recovered as $\Lambda \to 1$ (in this work we use a standard choice of $\Lambda=2.5$). The key feature of the Wilson chain is the behaviour of the hopping parameters $t_n$. For the metallic flat band, $t_n\sim \Lambda^{-n/2}$ at large $n$ \cite{Wilson1975}. This exponential energy-scale separation down the chain justifies a numerical scheme based on iterative diagonalization and truncation: starting from the dot, successive sites of the Wilson chain are coupled into the system and this intermediate Hamiltonian is diagonalized. Only the lowest $M_K$ eigenstates at iteration $n$ are used to construct the Hamiltonian at iteration $n+1$. High-energy states discarded at a given iteration do not affect the retained low-energy states at later iterations because of the ever-decreasing couplings $t_n$. This constitutes an RG procedure since the physics of the system at successively lower energy scales is revealed as more Wilson orbitals are added. The computational complexity is \textit{constant} as new Wilson orbitals are added (rather than exponentially growing) because the same number $M_K$ of states is kept at each step. Importantly, it was shown in Ref.~\cite{bulla1997anderson} that although the detailed structure of the Wilson chain coefficients are modified in the pseudogap DoS case, the energy scale separation down the chain is maintained, and hence the NRG can still be used in this case. We use the exact graphene DoS in this work rather than a pure pseudogap, and keep $M_K=6000$ states at each iteration. Dynamical quantities are calculated using the full-density-matrix NRG approach \cite{Weichselbaum2007,peters2006numerical}, established on the complete Anders-Schiller basis \cite{AndersSchiller2005}.


\section{Results and Discussion}
Having introduced the model and methods, we now discuss our NRG results in detail, starting with an overview of the phase diagram, RG flow diagram, and fixed point  analysis. In the following we confine our attention to the charge-degeneracy point $\delta V_g=0$. We also introduce the channel-asymmetry parameter $\Delta=J_2/J_1 \equiv W_2/W_1$ and discuss the physics in the space of $J\equiv J_1$, $W\equiv W_1$ and $\Delta$. Note that $\Delta=0$ corresponds to the situation in which channel $\alpha=2$ is decoupled on the level of the bare model, while for $0 < \Delta < 1$ both channels are coupled to the dot but channel $1$ couples more strongly.  $\Delta=1$ describes the frustrated two-channel situation. We confine our attention to regime $0\le \Delta \le 1$ but it should be noted that $\Delta>1$ simply corresponds to stronger coupling for channel $2$, and results follow from the duality $1 \leftrightarrow 2$ and $\Delta \leftrightarrow 1/\Delta$. We also assume $W>0$.


\subsection{Overview and phase diagram}\label{Sec:Overview}
\begin{figure*}
\hspace{-2cm} \begin{minipage}[b]{.4\textwidth}
\includegraphics[width=10cm]{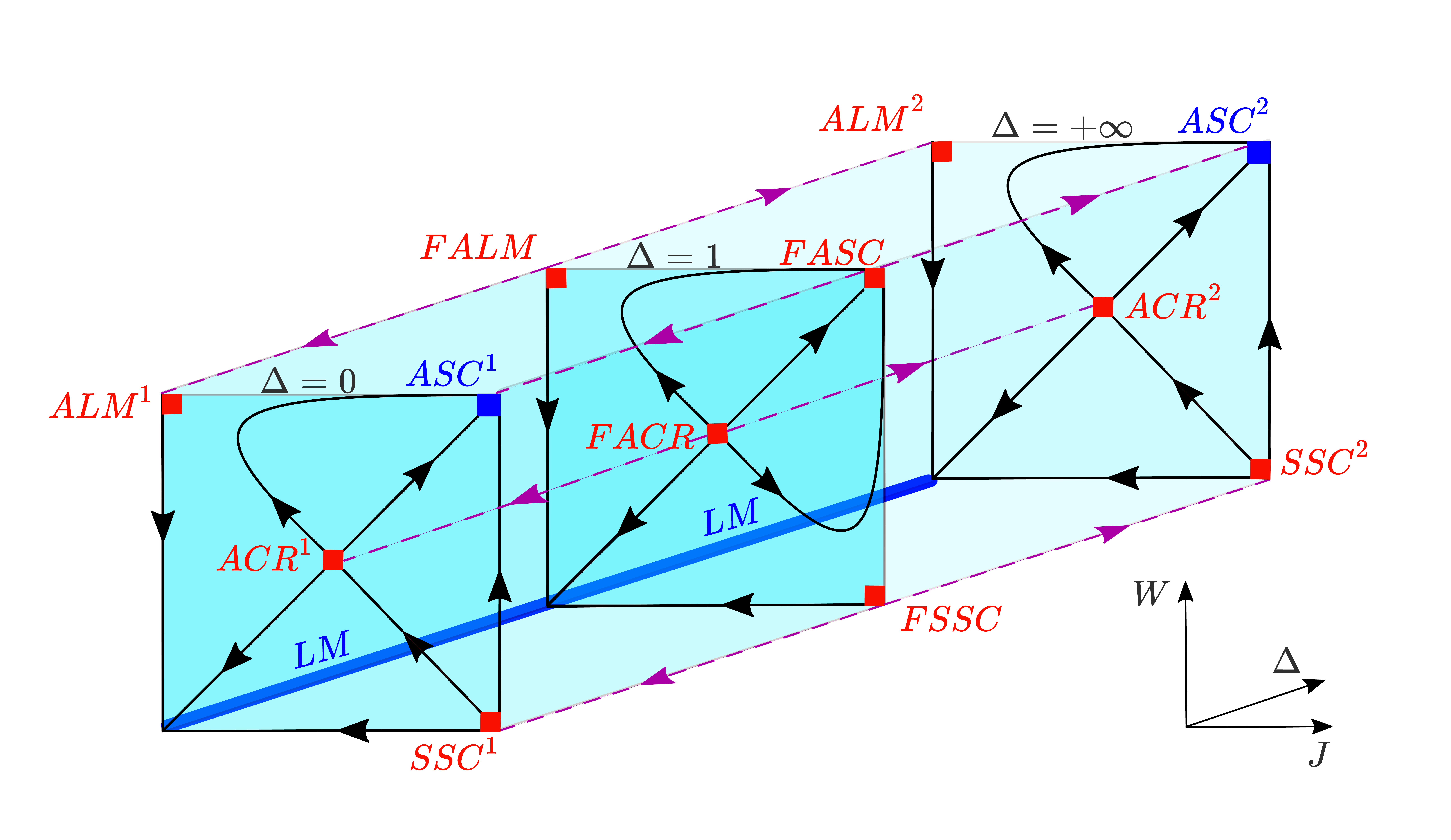}  \label{fig:RG}
\end{minipage}\qquad\qquad\qquad\qquad
\begin{minipage}[b]{.4\textwidth}
\includegraphics[width=9cm]{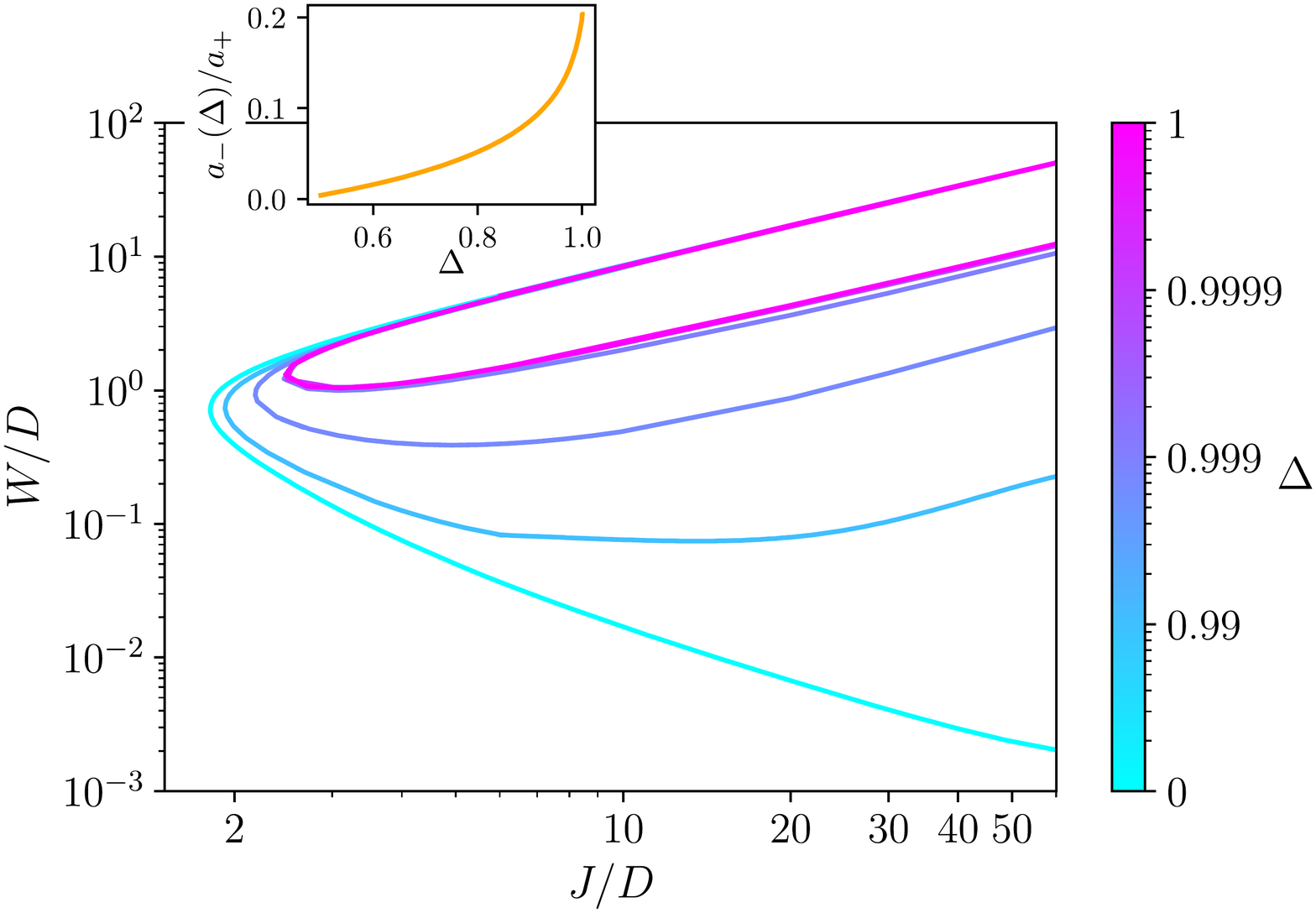}  \label{fig:PD}
\end{minipage}
\caption{\textit{Left:} RG flow diagram for the pseudogap 2CCK model Eq.~\ref{Eq:HK}, in the space of effective  exchange coupling $J$, potential scattering $W$ and channel asymmetry $\Delta$. Stable (unstable) FPs denoted as blue (red) squares. $\Delta=0$ ($1$) is the pure single-channel  (frustrated two-channel) model. For an explanation of the FPs, see text. \textit{Right:} Full NRG phase diagram for different $\Delta$. The enclosed region in each case is the Kondo-screened ASC phase (frustrated FASC for $\Delta=1$); the exterior region is the unscreened LM phase. Inset shows asymptotic behaviour of phase boundaries, see text.}\label{Fig:PH-RG}
\end{figure*}

The \textit{schematic} RG flow diagram in the space of $(J,W,\Delta)$ shown in the left panel of Fig.~\ref{Fig:PH-RG} was deduced from non-perturbative NRG results, and gives a good overview of the physics of Eq.~\ref{Eq:HK}. In the right panel we show the \textit{quantitative} phase diagram in the $(J,W)$ plane for different $\Delta$, with the exact phase boundaries obtained with NRG.

We first briefly recapitulate the results for the  single-channel (1CK) case obtained here for $\Delta=0$ (see Fig.~\ref{Fig:PH-RG}: front plane in RG diagram on the left, and turquoise line in the phase diagram on the right).  The basic physics are well-known from previous studies of the $r=1$ pseudogap Anderson and Kondo models \cite{bulla1997anderson,gonzalez1998renormalization,logan2000local,fritz2004phase,vojta2004upper,vojta2010gate,fritz2013physics}, although note that our graphene charge-Kondo setup gives a spin-anisotropic model, and we use the full graphene DoS rather than a pure pseudogap. At $W=0$ there is no Kondo effect: the symmetric strong coupling (SSC) FP is unstable and a finite potential scattering is required to screen the dot pseudospin. In this case the system flows to weak coupling and eventually to the free local moment (LM) FP with asymptotically decoupled conduction electrons. At finite $W$ an LM phase with an emergent particle-hole symmetry can be realized, in which the dot and leads decouple at low temperatures (that is $W, J\to 0$ under RG). However, at sufficiently strong bare $W$ the model supports a QPT to an asymmetric strong coupling (ASC) Kondo state, in which the dot pseudospin is screened and a single hole forms in the bath (that is $W, J \to \infty$ under RG). However the coupling $J$ must  overcome a critical threshold value (for Eq.~\ref{Eq:HK} $J_{CR}^{\rm min}\simeq 1.81D$). For $J < J_{CR}^{\rm min}$ no Kondo state is possible at any $W$. For $J \ge J_{CR}^{\rm min}$, the ASC FP is stable for $W^-_{CR}(J)< W < W^+_{CR}(J)$. In the lower branch we find $W^-_{CR}(J) \sim 1/J$ at large $J \gg J_{CR}^{\rm min}$, such that infinitesimal particle-hole symmetry breaking $W\to 0$ is required at large bare coupling $J\to \infty$. Thus, although SSC is unstable, the system can flow arbitrarily close to it, before ultimately crossing over to either ASC or LM. The transition between ASC and LM is first-order \cite{fritz2004phase,vojta2004upper}, and controlled by a particle-hole asymmetric critical FP denoted ACR. The full NRG phase boundary for our model at $\Delta=0$ is shown as the turquoise line in the right panel of Fig.~\ref{Fig:PH-RG}, and shows an interesting re-entrant behaviour back into the LM phase at large $W$ (we are not aware of a detailed discussion of this in the literature, even though the same behaviour arises in the regular spin-isotropic pseudogap Kondo model). This is physically intuitive since $J$ and $W$ work antagonistically: at very large $W$ the bath orbital $f_{1\sigma 0}$ becomes depopulated, and hence the exchange coupling to that site $J$ gets ``switched off''. Perturbative arguments suggest that the residual coupling to the $f_{1\sigma 1}$ bath orbital is then $J_{\rm eff}\sim t_0^2 J/W^2$, which is consistent with $W^+_{CR}(J) \sim J$ for the upper branch of the phase boundary. This is indeed confirmed by NRG calculations. 

\begin{table*}[t!]
\begin{center}
\caption{Classification of FPs according to their physical observables, with * denoting unstable FPs.\label{tabFP}}
\begin{tabular}{|c||c|c|c|c|c|c|}  
\hline

			~asymmetry~ & ~fixed-point~	& ~$S_{D}(T=0)$~	& ~$T\chi_{D}(T=0)$~ & ~$N_1$~ & ~$t_{1}(\omega,T\rightarrow0)$~ & ~$\mathcal{G}^{C}(\omega,T\rightarrow0)$~ \\
\hline\hline
$\forall \Delta$	& LM line & $\ln2$ & $1/4$ & $0$ & $|\omega|$ & $\omega^{2}$\\
\hline
$0 \leq \Delta<1$	& ASC	& $0$	& $0$ & $-1$  & $| \omega|$& $ \omega^{2}$\\
\hline

			$0 \leq \Delta <1$	& ACR$^{\star}$ & $\ln3$	& $1/6$	& ~~$-1/3$~~ & $1/\omega \ln^{2}(\lambda_{CR}/\omega)$ & $0$\\
\hline
			$\Delta=1$		&  FASC$^{\star}$	& $\ln2$	& $0$ & $-1/2$ & $|\omega|$ & $\omega^{2}$\\
			\hline

			$\Delta=1$		&  FACR$^{\star}$	& $\ln4$	& $1/8$ & $-1/4$ & ~~$1/\omega \ln^{2}(\lambda_{CR}/\omega)$~~ & $const$ \\
			\hline
		\end{tabular}
		\end{center}
\end{table*}

The main focus of this paper is the situation when the coupling to the second channel is switched on, $\Delta >0$, where we find several differences from the pure 1CK case. We discuss $0 < \Delta < 1$ first. Importantly, we find that the same LM and ASC phases are accessible, with $\Delta$ flowing to zero under RG flow on reducing the temperature or energy scale. Therefore, even though both channels are initially coupled to the dot (at high temperatures we have a free channel degree of freedom $\alpha=1,2$), any channel asymmetry leads asymptotically to the decoupling of the less strongly coupled channel $\alpha=2$ (this can be regarded as `channel freezing' at low temperatures). In the ASC phase, the dot flows to strong coupling with the more strongly coupled channel $\alpha=1$; while in the LM phase, channel $\alpha=1$ also eventually decouples, leaving a free dot pseudospin and free conduction electrons. This is indicated by the flow arrows towards the front plane in the RG diagram, Fig.~\ref{Fig:PH-RG} (left). 

However, at finite $\Delta$ the topology of the phase diagram changes -- see Fig.~\ref{Fig:PH-RG} (right). We still have a finite threshold value of the coupling to realize ASC physics, $J_{CR}^{\rm min}(\Delta)>0$  (which increases slightly from $\simeq 1.81D$ at $\Delta=0$ up to $\simeq 2.47D$ as $\Delta \to 1$). But the critical phase boundary now also develops a finite threshold value of the potential scattering $W_{CR}^{\rm min}(\Delta)>0$ (which reaches its maximum value $\simeq D$ as $\Delta \to 1$). Even at strong bare coupling $J\to \infty$ a finite $W$ is required to access the ASC phase. In fact, $W_{CR}^{\rm min}$ occurs at an intermediate value of $J$; at large $J$ we find $W^-_{CR}\sim J$ in the lower branch. For even larger $W$, we again have re-entrant LM behaviour, with an upper branch of the phase boundary. For $\Delta>0$ we therefore have large-$J$ behaviour $W^-_{CR}(J,\Delta)= a_{-}(\Delta)  J$ for the lower branch and $W^+_{CR}(J,\Delta)= a_{+}(\Delta)  J$ for the upper branch. Interestingly $a_{+}\approx 1$ independent of $\Delta$, while $a_{-}$ increases with increasing $\Delta$, as shown in the inset to the right panel of Fig.~\ref{Fig:PH-RG}. However, $a_{-}(\Delta)/a_{+}< 1$ for all $\Delta$ (the ratio reaches its maximum $\approx 0.2$ as $\Delta \to 1$) meaning that the upper and lower phase boundaries never cross, and the ASC phase persists out to infinite $J$ and $W$. The finite $W_{CR}^{\rm min}$ also implies that there is no crossover from SSC to ASC for $\Delta>0$.

Finally, we examine the channel-frustrated case $\Delta=1$ (see middle plane of Fig.~\ref{Fig:PH-RG}, left; and purple line of Fig.~\ref{Fig:PH-RG}, right). Here symmetry dictates a channel degeneracy down to $T=0$ and therefore no channel freezing. We find that the model supports an LM phase in which both channels flow symmetrically to weak coupling and to particle hole symmetry. However, the ASC FP is unstable since the Kondo effect and conduction electron hole in ASC occur in only one of the two channels. Instead we have a frustrated asymmetric strong coupling (FASC) phase, with a free channel degree of freedom (a doubled version of the ASC FP, with the Kondo effect and conduction electron hole forming in either channel $\alpha=1$ or $2$). The critical point separating LM and FASC in the $\Delta=1$ plane is denoted FACR. The frustrated FPs are delicate because they sit precisely on the separatrix between RG flow to states with dominant channel $1$ for $\Delta<1$, and flow to states with dominant channel $2$ for $\Delta>1$. Any finite perturbation $|1-\Delta|$ relieves the channel frustration and leads ultimately to channel freezing on the lowest energy scales. This QPT is also first-order; FACR is therefore tricritical since it sits at the boundary between LM, FASC and ASC.

In Table~\ref{tabFP} we summarize the FPs discussed above in relation to Fig.~\ref{Fig:PH-RG}, classifying them according to their physical properties. These properties are extracted from the limiting behaviour of the full thermodynamic and dynamic observables presented in the following.


\begin{figure*}[t!]
\centering
\includegraphics[width=18cm]{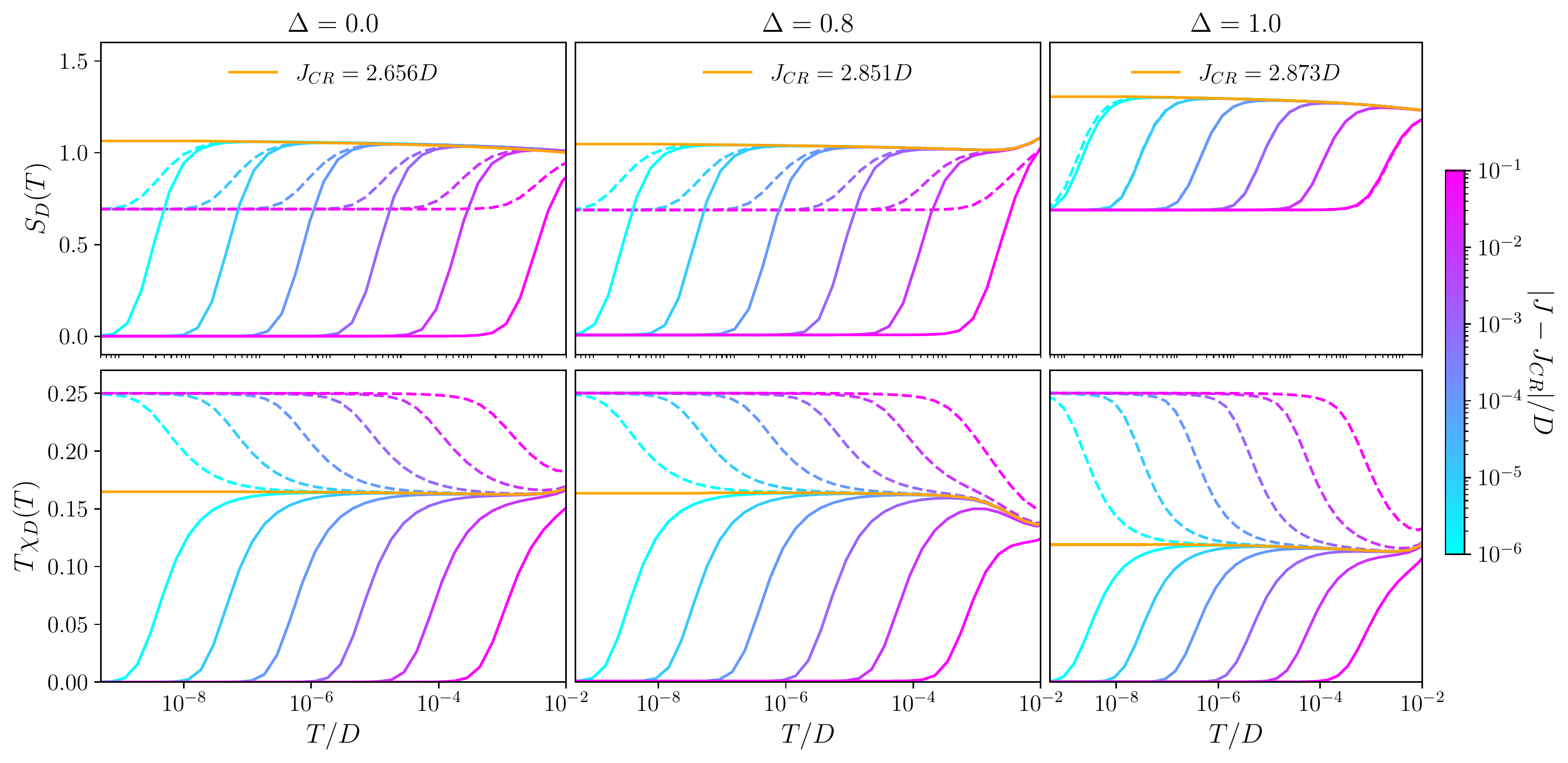}
\caption{Dot contribution to thermodynamic quantities for the graphene 2CCK model obtained by NRG. \textit{Top row:} entropy $S_D(T)$; \emph{bottom row:} magnetic susceptibility $T\chi_D(T)$. Left, middle, and right columns correspond to $\Delta=0$ (pure 1CCK), $\Delta=0.8$ (asymmetric 2CCK), and $\Delta=1$ (symmetric 2CCK) respectively. Shown for fixed $W=2D$, varying $J$ across the QPT according to the colour scale, with solid lines for $J>J_{CR}$ in the ASC (FASC) phase, and dashed lines for $J<J_{CR}$ in the LM phase. Orange lines show the behaviour at the ACR (FACR) critical point. 
\label{Fig:STChi}}
\end{figure*}   

\subsection{Thermodynamics and fixed points} \label{Sec:Thermo}
The temperature dependence of the dot contribution to entropy $S_D(T)$ and magnetic susceptibility $T\chi_D(T)$ are obtained from NRG \cite{Bulla2008} and presented in Fig.~\ref{Fig:STChi} for different channel asymmetries $\Delta$. We focus on the behaviour near the critical points by fixing $W$ and tuning $J$ across the transition. From this, information on the fixed points is deduced.


\subsubsection{Frozen channel degree of freedom: $0\leq \Delta<1$}
We first consider the regime $0\leq \Delta<1$ (left and middle columns of Fig.~\ref{Fig:STChi}). Solid lines show the behaviour in the ASC phase for $J>J_{CR}$, dashed lines for the LM phase with $J<J_{CR}$, and orange line at the critical point $J=J_{CR}$. 

In the LM phase, $S_D=\ln(2)$ and $T\chi_D=\tfrac{1}{4}$ at $T=0$ in all cases, characteristic of the asymptotically-free spin-$\tfrac{1}{2}$ dot pseudospin. The excess conduction electron charge (not shown) is zero in both channels, suggesting an emergent particle-hole symmetry. This is confirmed by analysis of the NRG many-particle level spectrum (finite size spectrum) at the LM FP, which is identical to that of the free leads. The dot remains unscreened in LM because of the depleted conduction electron DoS at low energies in graphene \cite{fritz2013physics}. The FP Hamiltonian in the LM phase is therefore given by,
\begin{equation}\label{eq:HLM}
\hat{H}_{LM}=\hat{H}_{2CCK} ~~ {\rm with} ~~ J_1=J_2=W_1=W_2=V_g=0
\end{equation}

In the ASC phase at $T=0$ we see quenched dot entropy $S_D=0$ and $T\chi_D=0$ in all cases, characteristic of Kondo singlet formation. However, the conduction electron excess charge is $N_1=-1$ and $N_2=0$ (for $W>0$), implying hole formation in the more strongly coupled channel $\alpha=1$ ($W_1 \to \infty$ under RG) while the less strongly coupled channel $\alpha=2$ recovers an effective low-energy particle-hole symmetry ($W_2 \to 0$ under RG). This suggests the screening mechanism in the generic two-channel case: as $W_1$ grows under RG, the $f_{1\sigma 0}$ Wilson orbital becomes depopulated, thereby generating an effective coupling between the dot pseudospin and the Wilson $f_{1\sigma 1}$ orbital, $J_{\textit{eff}}\sim t_0^2 J_1/W_1^2$. However, the DoS of the $f_{1\sigma 1}$ orbital is modified by the hole forming at the $n=0$ site. With $\rho(\omega)\equiv -\tfrac{1}{\pi}{\rm Im}\langle\langle f_{1\sigma 0} ; f_{1\sigma 0}^{\dagger}\rangle\rangle \sim |\omega|$ at low energies, we find $\rho_{\textit{eff}}(\omega)\equiv -\tfrac{1}{\pi}{\rm Im}\langle\langle f_{1\sigma 1} ; f_{1\sigma 1}^{\dagger}\rangle\rangle \sim 1/|\omega|$ at low energies. Therefore even though $J_{\textit{eff}}$ is perturbatively small, the effective DoS is strongly enhanced. The effective dimensionless RG flow parameter $j_1=\rho_{\textit{eff}} J_{\textit{eff}}$ grows under RG and leads to Kondo screening of the dot. The Kondo scale for this process is strongly enhanced because of the diverging effective DoS \cite{mitchell2013quantum}, and we find in practice that throughout the ASC phase $T_K \sim D$ (since $J_{CR}^{\rm min}>D$). 
However, no hole forms in the weakly coupled channel, and so $j_2=\rho J_2$ remains small due to the depleted bare DoS in channel 2, and flows under RG to weak coupling. This argument shows that the Kondo singlet must form in the same channel in which the hole forms. This is confirmed by analysis of the NRG level spectrum. 

In general we therefore have two distinct ASC phases and two distinct ASC fixed points, depending on whether $\Delta<1$ or $\Delta>1$. For $\Delta<1$, the hole-singlet complex forms in channel $\alpha=1$ and channel $\alpha=2$ decouples (FP denoted ASC$^1$), while for $\Delta>1$ (ASC$^2$) it is the other way around.  
The ASC$^{\alpha}$ FP Hamiltonian obtained when channel $\alpha$ is more strongly coupled reads,
\begin{equation}\label{Eq:HASC}
\begin{split}
\hat{H}^{\alpha}_{ASC} =& \hat{H}_{\rm leads} + J_{\alpha}\left (\hat{S}_D^+ f_{\alpha \downarrow 1}^{\dagger}f_{\alpha \uparrow 1}^{\phantom{\dagger}} + \hat{S}_D^- f_{\alpha \uparrow 1}^{\dagger}f_{\alpha \downarrow 1}^{\phantom{\dagger}}\right ) \\ &+ W_{\alpha}\sum_{\sigma}f_{\alpha \sigma 0}^{\dagger}f_{\alpha \sigma 0}^{\phantom{\dagger}} \qquad {\rm with} \qquad J_{\alpha},W_{\alpha}\to \infty \;.
\end{split}
\end{equation}

We now consider the situation in the close vicinity of the QPT, by fixing $W$ and tuning $J$. At the critical point itself (orange line, $J=J_{CR}$) we find $S_D=\ln(3)$ and $T\chi_D=\tfrac{1}{6}$ at $T=0$. This suggests a level-crossing transition in which the critical FP ACR comprises uncoupled sectors corresponding to LM and ASC. This gives an overall dot ground state degeneracy of $2+1=3$ states (2 for LM, 1 for ASC) consistent with the $\ln(3)$ entropy, and a magnetic susceptibility $(\tfrac{1}{4}+\tfrac{1}{4}+0)/3=\tfrac{1}{6}$ (corresponding to the average of $(S^z)^2$ for these three degenerate states). This is further supported by the conduction electron excess charge $N_1=-\tfrac{1}{3}$, since a single hole appears in channel 1 for only one of the three degenerate ground states (and $N_2=0$ for the decoupled free channel 2). The first-order transition is also consistent with the \textit{linear} crossover scale $T^* \sim |J-J_{CR}|$, describing the flow from ACR to either LM or ASC due to a small detuning perturbation. This scale is evident in Fig.~\ref{Fig:STChi} by the sequence of lines for different  $(J-J_{CR})$. Indeed, one can cross the QPT by fixing $J$ and tuning $W$ through $W_{CR}$, which also gives a linear scale $T^*$. We also checked this behaviour along the entire critical phase boundary lines $(J_{CR},W_{CR})$ in Fig.~\ref{Fig:PH-RG} for different $\Delta$. We find,
\begin{eqnarray}\label{eq:tstar}
T^* = b|J-J_{CR}| + c|W-W_{CR}| \;,
\end{eqnarray}
where $b\equiv b(J_{CR},W_{CR},\Delta)$ and $c\equiv c(J_{CR},W_{CR},\Delta)$. This implies a universal scaling in terms of a single reduced parameter $T/T^*$, independent of the combination of bare perturbations that act.
The FP Hamiltonian describing the critical point is,
\begin{equation}\label{Eq:HACR}
\hat{H}^{\alpha}_{ACR} = \left(\frac{1+\hat{\tau}^{z}}{2}\right)\hat{H}_{LM} + \left(\frac{1-\hat{\tau}^{z}}{2}\right)\hat{H}^{\alpha}_{ASC}~.
\end{equation}
where the $\alpha$ label denotes the more strongly coupled lead with which the dot forms the Kondo effect in ASC, and $\hat{\tau}^z$ is a Pauli-$z$ operator. In Eq.~\ref{Eq:HACR}, $\tau^z=+1$ gives the doubly-degenerate LM ground state ($\hat{H}_{LM}$ given in Eq.~\ref{eq:HLM}) while $\tau^z=-1$ gives the ASC ground state ($\hat{H}_{ASC}$ given in Eq.~\ref{Eq:HASC}). At the ACR FP, the three many-body ground states are degenerate and uncoupled ($\hat{\tau}$ has no dynamics). Since ACR is unstable, we also consider the leading RG relevant perturbations to the FP Hamiltonian, $\delta \hat{H}^{\alpha}_{ACR} \sim T^*\hat{\tau}^z$, which has the effect of biasing towards either the LM or ASC ground states on the scale of $T^*$.

The qualitative behaviour of the thermodynamics shown in Fig.~\ref{Fig:STChi} for $\Delta=0$ and $\Delta=0.8$ is similar, but it should be noted that both channels are involved for $\Delta \ne 0$ at finite temperatures. However, the less strongly coupled channel decouples asymptotically because finite $0<\Delta<1$ flows to $\Delta=0$ under RG on reducing the temperature or energy scale (see Fig.~\ref{Fig:PH-RG}, left).


\subsubsection{Frustrated channel degree of freedom: $\Delta=1$}
We turn now to the frustrated case $\Delta=1$, with pristine channel symmetry -- see Fig.~\ref{Fig:STChi}, right column. Although the $T=0$ entropy is $S_D=\ln(2)$ everywhere except on the phase boundary (top right panel of Fig.~\ref{Fig:STChi}), the origin of the ground state degeneracy is different in the two phases separated by it. In the LM phase (realized for $J<J_{CR}$), we again have a free dot pseudospin decoupled from two symmetric baths of free conduction electrons; the $\ln(2)$ entropy here derives from the free dot pseudospin-$\tfrac{1}{2}$ degree of freedom. This is confirmed by the magnetic susceptibility in this phase, which reaches $T\chi_D=\tfrac{1}{4}$ (dashed lines, bottom right panel of Fig.~\ref{Fig:STChi}). The other phase (realized for $J>J_{CR}$) is described by the FASC FP: due to the channel symmetry, the ASC state can form in either channel $\alpha=1,2$.  The $\ln(2)$ entropy in this case  derives from the free channel degree of freedom \cite{schneider2011two}, which embodies the choice of forming the hole-singlet complex of ASC with either of the two channels. This is reflected in the $T=0$ value of $T\chi_D=0$ in the FASC phase (solid lines, bottom right panel of Fig.~\ref{Fig:STChi}), since the dot pseudospin is Kondo screened in both of the degenerate ground states. Furthermore we find that the average conduction electron excess charge in FASC is $N_{\alpha}=\tfrac{1}{2}$ for both channels -- that is, a single hole forms, with equal probability to be in either channel 1 or 2.

A Kondo strong coupling state involving both channels simultaneously is not stable. To see this, consider two holes forming symmetrically in the $f_{1\sigma 0}$ and $f_{2\sigma 0}$ Wilson $n=0$ orbitals ($W_1=W_2 \to \infty$), and effective Kondo couplings $J_{1,\textit{eff}}=J_{2,\textit{eff}}\to \infty$ between the dot pseudospin and the residual Wilson $n=1$ orbitals $f_{1\sigma 1}$ and $f_{2\sigma 1}$, which have an effective DoS $\rho_{\textit{eff}}(\omega)\sim 1/|\omega|$ -- a channel symmetric version of the usual hole-singlet mechanism in ASC as described in the previous section. However, the dot entropy is not quenched in this case, since the ground state of the complex is a spin-doublet. This effective doublet state couples to the Wilson $n=2$ orbitals $f_{1\sigma 2}$ and $f_{2\sigma 2}$. But since the DoS of these sites is again $\sim |\omega|$ the effective local moment cannot be screened and the system flows to the LM FP. The dot pseudospin can only be screened by an \textit{asymmetric} ASC state. Channel symmetry is restored by having two such degenerate states, one in each channel. 

The FASC FP Hamiltonian comprises a combination of $\hat{H}_{ASC}^1$ and $\hat{H}_{ASC}^2$ from Eq.~\ref{Eq:HASC}, controlled by an emergent channel degree of freedom $\hat{\alpha}$,
\begin{equation}\label{Eq:HFASC}
\hat{H}_{FASC} = \left(\frac{1+\hat{\alpha}^{z}}{2}\right)\hat{H}^{1}_{ASC}+\left(\frac{1-\hat{\alpha}^{z}}{2}\right)\hat{H}^{2}_{ASC}~.
\end{equation}
Here $\hat{\alpha}^z$ is a Pauli-$z$ operator that selects ASC$^1$ when $\alpha^z=+1$ and ASC$^2$ when $\alpha^z=-1$. Restricting to the symmetric $\Delta=1$ plane, FASC is stable. However, there is an instability with respect to breaking channel symmetry (not shown), since then either ASC$^1$ or ASC$^2$ will be selected on the lowest energy scales. A finite perturbation $|1-\Delta|$ generates a flow from FASC to ASC$^1$ or ASC$^2$; from NRG we find that this QPT is also first-order. The low-energy scale determining the crossover is $T_{\Delta} \sim |1-\Delta|$. This can be captured in the effective model by including the leading RG relevant perturbation to the FASC FP, $\delta \hat{H}_{FASC} \sim T_{\Delta} \hat{\alpha}^z$.

Finally, we consider the quantum critical point in the $\Delta=1$ plane between LM and FASC. Here we find a level crossing (first-order) transition, with entropy $S_D=\ln(4)$ and magnetic susceptibility $T\chi_D=\tfrac{1}{8}$ at the FACR FP \cite{schneider2011two}, which derives from the composition of uncoupled LM and FASC sectors. We have two spin-$\tfrac{1}{2}$ states from the LM degenerate with two spin-singlet states with a free channel degree of freedom in FASC. The excess conduction electron charge is therefore $N_{\alpha}=-\tfrac{1}{4}$ per channel. We describe the FACR FP with the Hamiltonian,
\begin{equation}\label{Eq:HFACR}
\hat{H}_{FACR} = \left(\frac{1+\hat{\tau}^{z}}{2}\right)\hat{H}_{LM}+\left(\frac{1-\hat{\tau}^{z}}{2}\right)\hat{H}_{FASC} ~,
\end{equation}
where $\hat{H}_{LM}$ is given in Eq.~\ref{eq:HLM} and $\hat{H}_{FASC}$ in Eq.~\ref{Eq:HFASC}, and we have introduced the operator $\hat{\tau}^z$ to distinguish the sectors, similar to Eq.~\ref{Eq:HACR}. As with ACR, the FP is destabilized by RG relevant detuning perturbations that favour either LM of FASC, which collectively generate the scale $T^*$ given in Eq.~\ref{eq:tstar}. This leads to an FP correction $\delta \hat{H}_{FACR}^*\sim T^* \hat{\tau}^z$. This is shown by the sequence of lines in the right column of Fig.~\ref{Fig:STChi}. However, FACR is also destabilized by relieving the channel frustration through the perturbation $|1-\Delta|$ which generates the scale $T_{\Delta}$, since FACR contains an FASC sector with this instability. Therefore FACR has a second RG relevant correction 
$\delta \hat{H}_{FACR}^{\Delta}\sim T_{\Delta} \hat{\alpha}^z$. FACR is in this sense \textit{tricritical} since it sits between LM, FASC and ASC.

\begin{figure*}[t!]
\centering
\includegraphics[width=14.7cm]{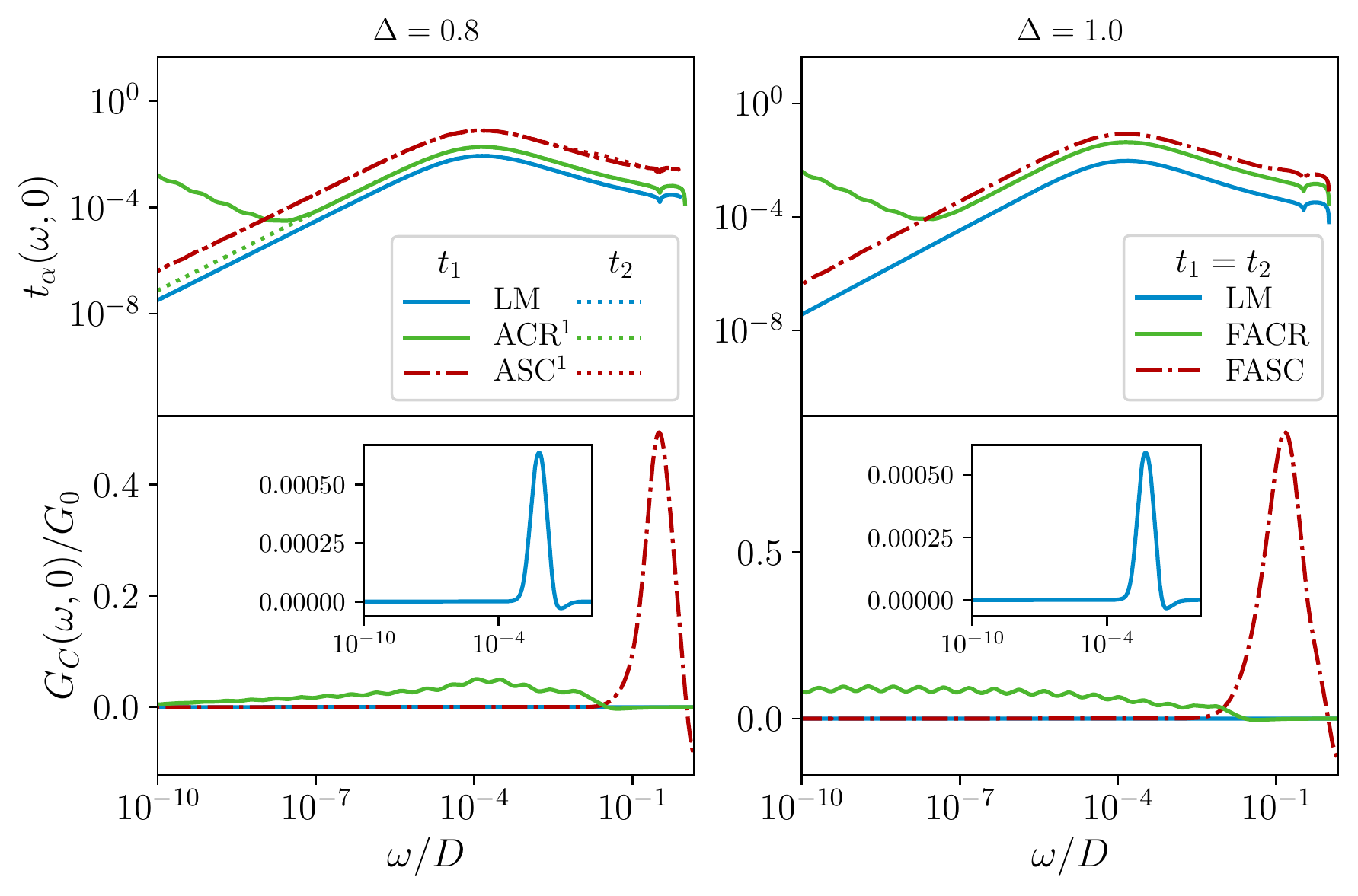}
\caption{NRG results for dynamics and transport in the graphene 2CCK model at $T=0$. \textit{Top row:} Channel-resolved spectrum of the $\mathrm{T}$-matrix $t_{\alpha}(\omega,0)$. \textit{Bottom row:} Linear response ac electrical conductance $G_C(\omega,0)$. \textit{Left}: channel asymmetry $\Delta=0.8$; \textit{Right} frustrated case $\Delta=1$. Model parameters: $J/D=10$ with $W/D=1$ (LM); $12$ (ASC); $12$ (FASC); $\simeq 8.605$ (ACR); $\simeq 8.572$ (FACR).
\label{Fig:TGc}}
\end{figure*}   

\subsection{Dynamics and transport}\label{Sec:TmatGc}
We now discuss the low-temperature behaviour of the scattering T-matrix and linear response ac electrical conductance in the graphene 2CCK device -- see Fig.~\ref{Fig:TGc}. 
We first consider the $T=0$ spectrum of the T-matrix as a function of energy in the top row of Fig.~\ref{Fig:TGc}, for the channel asymmetric case $\Delta=0.8$ (left) and frustrated case $\Delta=1$ (right). In all cases we identify an emergent low-energy scale $\lambda$ (which is $\approx 10^{-4}D$ for the parameters chosen) which characterizes the RG flow through a  crossover behaviour in the pseudogap dynamics \cite{fritz2004phase,schneider2011two}. 

Deep in the LM phase (blue solid and dotted lines), the bare potential scattering $W$ modifies the bare conduction electron pseudogap DoS of graphene $\rho(\omega)\sim |\omega|$, to give an effective DoS $\rho_{\textit{eff}}(\omega)\sim 1/|\omega|$ ($\pi/2$ phase shift) up to logarithmic corrections. This produces leading behaviour in the T-matrix $t_{\alpha}(\omega,0) \sim 1/|\omega|$, as seen in Fig.~\ref{Fig:TGc} for $|\omega| \gg \lambda$. However, under RG $W \to 0$ in the LM phase; this flow is controlled by the scale $\lambda$. Therefore, on the scale of $\lambda$ the effective DoS returns to $\sim |\omega|$ (zero phase shift) and hence $t_{\alpha}(\omega,0) \sim |\omega|$ for $|\omega|\ll \lambda$. Since the emergent particle-hole symmetry in the LM phase occurs in both channels for any $\Delta$, we see the same behaviour for $t_1(\omega,0)$ and $t_2(\omega,0)$ for both $\Delta=0.8$ and $1$. 

In the ASC phase for $\Delta=0.8$, the weakly coupled channel $\alpha=2$ shows the same behaviour as LM since it decouples from the dot and gains particle-hole symmetry. For the strongly coupled channel $\alpha=1$, we have the hole-singlet mechanism in which both the effective $W,J \to \infty$. Counterintuitively we again see similar dynamical behaviour as for LM. This is because, for $|\omega| \gg \lambda$ we have a developing conduction electron hole which gives $t_{1}(\omega,0) \sim 1/|\omega|$, while for $|\omega|\ll \lambda$ the Kondo singlet forming with the $n=1$ Wilson orbital effectively removes a second site from the bath. The remaining conduction electrons experience a $\pi$ phase shift from the modified boundary, and the effective DoS is back to $\sim |\omega|$. Therefore in ASC we also have $t_1(\omega,0) \sim |\omega|$ for $|\omega|\ll \lambda$. That we have identical behaviour for $\Delta=1$ (in both channels) confirms that FASC is indeed a superposition of ASC$^1$ and ASC$^2$ as argued above.

On the lowest energy scales we have $t_{\alpha}(\omega,0)\sim|\omega|$ in both channels, at any $\Delta$, and in either phase. Given the bare DoS $\rho(\omega)\sim |\omega|$, this confirms that both phases are regular Fermi liquids, with well-defined (long-lived) quasiparticles \cite{Hewson,fritz2004phase,vojta2004upper}.

More interesting is the behaviour at the critical point (F)ACR, since here we have both spin and charge fluctuations associated with the degenerate LM and (F)ASC ground states. A new dynamical scale is generated, $\lambda_{CR}\sim \lambda^2/D$ ($\approx 10^{-8}D$ for the chosen parameters) which characterizes the low-energy RG flow \cite{fritz2004phase,schneider2011two}. We find from NRG that in the channel-asymmetric graphene 2CCK model (e.g.~at $\Delta=0.8$ as shown) the T-matrix of the more strongly-coupled channel $\alpha=1$ \textit{diverges} at low energies. Specifically, $t_1(\omega ,0)\sim 1/[|\omega/\lambda_{CR}|\times\ln^{2}(|\omega/\lambda_{CR}|)]$ as $|\omega|\to 0$ (solid green lines in the top panels of Fig.~\ref{Fig:TGc}) \cite{vojta2001kondo}, indicating that ACR is an NFL FP. The dynamical crossover, and hence the minimum in $t_1(\omega,0)$ occurs on the scale of $|\omega|\propto \lambda_{CR}$. However, the weakly coupled channel $\alpha=2$ has FL correlations $t_2(\omega ,0)\sim |\omega|$ as $|\omega|\to 0$ (dotted green lines), confirming that it decouples from the critical complex formed from the dot and channel 1. In the frustrated case $\Delta=1$, both channels behave identically -- and both exhibit the same NFL critical divergence at low energies. This again suggests that FACR comprises two copies of ACR, one in each channel. The enhanced conduction electron scattering at the critical point has implications for the conductance, as now shown.

In the bottom row of Fig.~\ref{Fig:TGc} we plot the $T=0$ dynamical ac conductance as a function of ac driving frequency $\omega$, for the same set of systems. The dc conductance is obtained in the $\omega\to 0$ limit, which we consider first. In the charge-Kondo system, series transport proceeds by the following mechanism: an electron tunnels from the source lead onto the dot (say at QPC $\alpha=1$), thus flipping the dot charge pseudospin from $\Downarrow$ to $\Uparrow$. A second electron then tunnels from the dot to the drain lead (at QPC $\alpha=2$), thus flipping the dot charge pseudospin back to $\Downarrow$ and ``resetting'' the device ready for transport of another electron. A bias voltage between source and drain produces a net current flow. The amplitude for such a process depends on the conduction electron density of states $\rho(\omega)$ and the tunneling rate at the QPCs. For graphene we have $\rho(\omega)\sim |\omega|$ at low energies, suggesting that the low-temperature dc conductance should vanish, since there are not enough low-energy electrons in the graphene leads to tunnel through the nanostructure. On the other hand, the tunneling rate gets renormalized by the interactions (the energy-dependent scattering at the QPCs is characterized by the T-matrices discussed above). Indeed, strong renormalization of the bare QPC transmission at low temperatures due to Kondo physics was measured experimentally in the metallic leads version of the present system in Ref.~\cite{2CK-RenormalisationFlow_IftikharPierre2015Exp}. 

The measured dc conductance of the graphene 2CCK involves a subtle interplay between the conduction electron DoS and interaction-renormalized scattering rates. We expect the $T=0$ dc conductance to vanish in all channel-asymmetric systems because the less strongly coupled channel always decouples on the lowest energy scales. Both leads must remain coupled to ensure a finite series current. This is indeed seen in the $\omega \to 0$ limit of each of the curves in the bottom left panel of Fig.~\ref{Fig:TGc} for $\Delta=0.8$. 
However, in the frustrated (channel-symmetric) case $\Delta=1$, both channels remain coupled down to $T=0$. Although the scattering rates and bare DoS both vanish as $\sim |\omega|$ in the LM and FASC phases implying a suppression of dc conductance, at the critical point FACR the electronic scattering diverges as $|\omega|\to 0$. We find from NRG that these effects conspire to give a \textit{finite} linear dc conductance in this case -- see green line in the bottom right panel of Fig.~\ref{Fig:TGc}.

For an ac bias the conductance is measured as a function of the driving frequency $\omega$. Conductance resonances are expected when the ac frequency matches the QPC tunneling rate. At high energies the pseudospin flip rate in the 2CCK model is given by the bare $J$ (or effective $J_{\text{eff}}$). We therefore expect to see a peak in the ac conductance when $|\omega|\sim J,J_{\textit{eff}}$; this is observed from NRG results in the LM, ASC and FASC phases in Fig.~\ref{Fig:TGc}. However, at low energies $|\omega|\ll \lambda$, the pseudospin flip rate is renormalized and we find $G_C(\omega,0)\sim \omega^2$ in these cases, independent of $\Delta$. At the critical point ACR for $0< \Delta <1$, both charge and spin fluctuations give an enhanced ac conductance around $|\omega|\sim \lambda$. However, channel $\alpha=2$ decouples for $|\omega|\ll \lambda$ and so the conductance also decays at low frequencies. We find from NRG a slow attenuation $G_C(\omega,0) \sim -1/\ln|\omega|$ in this regime. However, in the channel-symmetric case $\Delta=1$ at the critical point FACR, $G_C(\omega,0) \sim \textit{const}.$ for $|\omega|\ll \lambda_{CR}$. The finite dynamical conductance here persists down to the dc limit. This is the smoking gun signature of the NFL frustrated critical point in the graphene 2CCK system.


\section{Conclusion and Outlook}\label{Sec:Discus}
In this paper we propose a charge-Kondo quantum dot device made from graphene components that realizes a linear-pseudogap two-channel Kondo model. This exotic system has a complex phase diagram in the space of dot-lead coupling strength, potential scattering, and channel asymmetry. We analyze the thermodynamic and dynamical properties of the model using NRG, and use this to gain a detailed understanding of the renormalization group flow and fixed points. We classify the fixed points and construct the corresponding fixed-point Hamiltonians, together with their leading corrections. In particular, we uncover a channel-frustrated Kondo phase, with a non-Fermi liquid quantum critical point at the first-order quantum phase transition. Despite the depleted electronic density of the neutral graphene leads at low energies, critical fluctuations give rise to diverging scattering rates at the critical point, which produce a finite conductance even as $T\to 0$.

The model supports other interesting but as yet unexplored regimes. We have confined attention to the dot charge degeneracy point $\delta V_g=0$; but finite $\delta V_g$ appears in the effective model like a magnetic field on the dot pseudospin. One could also investigate the effect of doping/gating the graphene so that the Fermi level is not at the Dirac point. This will give rise to a quantum phase transition between metallic 2CK and pseudogap 2CK, with the competition controlled by the ratio of the doping to the Kondo temperature. Other physical quantities could also be investigated, such as thermoelectric transport with a temperature gradient between leads.

The pseudogap 2CCK model we have studied theoretically here is likely a simplified description of any real graphene charge-Kondo nanoelectronics device. There may be complexities and subtleties in an experimental realization which are not included in our model or analysis. For example, we have assumed that the conduction electrons on both leads and dot have the same DoS. In particular, gate voltage tuning of the dot to achieve charge-degeneracy, and the addition of the decoherer, may affect the dot electronic DoS. However, we do not expect our basic results to be qualitatively modified by this because the Kondo exchange interaction derives from tunneling at the QPC and hence involves the DoS of both lead and dot (see expression for $\tau(\omega)$). Therefore even if only the lead DoS is pseudogapped at low energies, an effective pseudogap Kondo model should still result. To make quantitative connection to experiment, such effects would have to be taken into account, as well as the possible involvement of more than just 2 dot charge states (that is, relaxing the condition $T\ll E_C$). We believe the predicted conductance signature of the frustrated quantum critical point should however still be observable in experiment.


\begin{acknowledgments}
We acknowledge funding from the Irish Research Council through the Laureate Award 2017/2018 grant IRCLA/2017/169 (AKM/JBR) and the Enterprise Partnership Scheme grant EPSPG/2017/343 (ELM)
\end{acknowledgments}

\section*{Abbreviations}
The following abbreviations are used in this manuscript:\\
\noindent 
\begin{center}
\begin{tabular}{@{}ll}
QD & quantum dot\\
QPC & quantum point contact\\
QPT & quantum phase transition\\
QCP & quantum critical point\\
2CK & two channel Kondo\\
NRG & numerical renormalization group  \\
FL & Fermi liquid\\
NFL & non-Fermi liquid\\
DoS & density of states \\
RG & renormalization group \\
FP & fixed-point \\
LM & local moment \\
(F)ALM & (frustrated) asymmetric local moment \\
(F)ASC & (frustrated) asymmetric strong coupling \\
(F)SSC & (frustrated) symmetric strong coupling \\
(F)ACR & (frustrated) asymmetric critical \\
\end{tabular}
\end{center}
\vspace{20pt}


\bibliography{biblio}

\end{document}